# A framework for rapid, reproducible, and high-fidelity whole-brain multi-pool CEST imaging at 3T

Yupeng Wu [a], Siyuan Fang [a], Siyuan Wang [b], Caixia Fu [c], Jianqi Li [a,d,e,*]

[a] Shanghai Key Laboratory of Magnetic Resonance, School of Physics and Electronic Science, East China Normal University, Shanghai, China

[b] Zhejiang A&F University, Hangzhou, China

[c] MR Collaboration, Siemens (Shenzhen) Magnetic Resonance, Shenzhen, China

[d] Institute of Magnetic Resonance and Molecular Imaging in Medicine, East China Normal University, Shanghai, China

[e] Institute of Brain and Education Innovation, East China Normal University, Shanghai, China

***Corresponding author:**

Jianqi Li, Ph.D, Shanghai Key Laboratory of Magnetic Resonance, School of Physics and Electronic Science, East China Normal University, 3663 North Zhongshan Road, Shanghai, 200062, China; Email: jqli@phy.ecnu.edu.cn.

## Abstract

**Purpose:** To develop and validate a framework for rapid, accurate, and reproducible whole-brain, multi-pool chemical exchange saturation transfer (CEST) imaging at 3T, addressing challenges of long acquisition times and confounding factors.

**Methods:** A single-shot 3D true fast imaging with steady-state precession (True FISP) sequence was optimized for whole-brain multi-pool CEST. Rapid $B_0$, $B_1$, and $T_1$ mapping was performed using a dual-echo modified four-angle method. A feed-forward neural network was developed for rapid $B_1$ correction, trained against the conventional multi-power method. The apparent exchange-dependent relaxation (AREX) metric was used to correct for $T_1$ and magnetization transfer (MT) effects. The framework was validated in phantoms and healthy human subjects (N=8), including a test-retest reproducibility assessment.

**Results:** The True FISP sequence yielded high-quality, whole-brain images with minimal artifacts and distortion in a clinically feasible scan time (~9 minutes). Phantom studies confirmed the effectiveness of $B_1$ correction (coefficient of variation [CV] for $MT_MTR_{LD}$ decreased from 22.49% to 4.61%) and AREX-based confounder correction (CV for APT_AREX reduced from 33.6% to 6.9%). The neural network $B_1$ correction showed excellent agreement with the conventional multi-power method in vivo (ICC > 0.97). High test-retest reproducibility was demonstrated across 96 brain regions, with the average CV for APT_AREX under 10% for over 95% of regions.

**Conclusion:** A rapid and robust framework for whole-brain quantitative multi-pool CEST imaging was successfully developed and validated. By integrating an efficient acquisition sequence with a streamlined correction pipeline, this approach overcomes key barriers to clinical translation, enabling reliable metabolic imaging for widespread brain pathologies.

## 1 INTRODUCTION

Chemical exchange saturation transfer (CEST) is a powerful molecular imaging technique for non-invasive detection of molecules and physiological parameters, such as proteins and pH levels. This capability makes the technique valuable in clinical research for applications in oncology[1-4], neurology[5-7], and neurodegenerative diseases[8-12]. A key advancement of this technique is multi-pool CEST, which acquires data at multiple frequency offsets to simultaneously provide information from several molecular sources[1]. The multi-pool approach provides a richer characterization of the tissue's metabolic state, enhancing diagnostic specificity for complex pathologies[9,10]. While this method enhances diagnostic efficiency and improves the interpretability of imaging results[8-11,13], it inherently requires longer acquisition times compared to single-pool methods.

This challenge is magnified by the clinical need for whole-brain coverage. For diseases that affect multiple brain regions and gray matter nuclei, such as neurodegenerative disorders, whole-brain imaging is particularly advantageous as it facilitates comprehensive, region-of-interest-based analysis. For example, in the study of Alzheimer's and Parkinson's disease, the ability to reliably quantify metabolic changes simultaneously across critical structures—such as the hippocampus, basal ganglia, and various cortical areas—is of paramount importance[13]. However, combining the time-intensive nature of multi-pool analysis with the demands of whole-brain acquisition creates a significant bottleneck for clinical translation. Consequently, most clinical multi-pool CEST studies have been restricted to single-slice or partial brain acquisitions[4,9,14], which severely limits the technique's practical utility for studying widespread brain pathologies.

Beyond the challenge of long acquisition times in whole-brain multi-pool CEST, the accuracy of quantitative CEST is further complicated by several confounding factors. A primary concern is the hardware-related magnetic field inhomogeneities, namely the main magnetic field ($B_0$) and the transmit radio-frequency field ($B_1$). Although $B_0$ field inhomogeneity can distort the spectrum, its correction is relatively straightforward. In contrast, correcting for $B_1$ field inhomogeneity presents a significant challenge, as conventional methods require acquiring the Z-spectrum at multiple power levels, which further exacerbates the already long scan time[15,16]. Furthermore, ensuring accurate biological interpretation remains a challenge. Although multi-pool modeling can separate spectrally overlapping signals, the magnitude of each resolved component is still suppressed by broad background effects. These are primarily the powerful direct water saturation (often termed 'spillover') and the underlying semi-solid magnetization transfer (MT) effect, both of which dilute the specific CEST signal[17,18]. Additionally, this signal is scaled by the tissue's $T_1$ relaxation time[17]. To correct for these confounding background effects, advanced metrics such as the apparent exchange-dependent relaxation (AREX)—which is based on an inverse Z-spectrum analysis—have been proposed to remove the influence of both spillover and the underlying MT dilution, as well as $T_1$ effects[1,18]. However, implementing AREX requires acquiring dedicated $T_1$ maps, adding to the total scan time.

Among various efforts to accelerate CEST imaging, our recent work[19] demonstrated that a single-shot 3D multi-pool CEST sequence based on a true fast imaging with steady-state precession (True FISP) readout offers a pathway to higher signal acquisition efficiency. This sequence provides a superior signal-to-noise ratio (SNR) compared to conventional spoiled GRE readouts within the same acquisition time, establishing a promising foundation for clinical translation. However, that initial work was limited to partial brain coverage and did not integrate a comprehensive framework for correcting the other critical confounding factors, such as field inhomogeneities and $T_1$/MT effects.

Building on that promising foundation, this study’s primary goal is to optimize this True FISP-based approach to achieve true whole-brain coverage and to develop a comprehensive solution for clinically applicable, whole-brain, multi-pool CEST imaging. To achieve this, we focused on the following objectives: (1) to optimize and validate a whole-brain multi-pool CEST sequence based on the True FISP readout; (2) to develop rapid acquisition sequences for $B_0$, $B_1$, and $T_1$ maps and to validate the effectiveness of $B_1$ correction and the AREX metric in phantoms and human subjects; (3) to integrate and validate a neural network model[20] for rapid $B_1$ correction, a technique that has shown promise for reducing scan times, and to validate its performance in human subjects; and (4) to verify the region-level reproducibility of the derived CEST metrics in healthy subjects.

## 2 METHODS

### 2.1 Study design and participants

#### 2.1.1 Study overview

The study was structured in four parts to systematically develop and validate a comprehensive framework for whole-brain multi-pool CEST imaging. Study 1 aimed to develop a whole-brain multi-pool CEST protocol based on a True FISP readout and demonstrate its image quality. Study 2 evaluated the effectiveness of the traditional three-point $B_1$ correction method[16] and the AREX metric in phantoms and human subjects. Study 3 developed and validated a neural network for rapid $B_1$ correction in human subjects. Finally, Study 4 assessed the region-level test-retest reproducibility of the entire pipeline in a separate cohort of healthy volunteers. All procedures were approved by the local institutional review board, and written informed consent was obtained from all participants.

#### 2.1.2 Phantom preparation

Two custom-built phantoms were used for validation experiments in Study 2. The first phantom was designed to validate the traditional three-point $B_1$ correction method and to characterize the effect of $B_1$ inhomogeneity on CEST measurements. It consisted of six identical tubes, each containing a solution of 8% polylysine and 1% agarose, adjusted to pH 7.3. The second phantom was designed to evaluate the efficacy of the AREX metric in correcting for $T_1$ and MT confounders. It also contained six tubes, all maintained at pH 7.3, with the following compositions: Tube 1, 8% polylysine with 1× gadolinium contrast agent (0.0625 mM); Tube 2, 8% polylysine; Tube 3, 8% polylysine with 1% agarose; Tube 4, 8% polylysine with 2× gadolinium contrast agent (0.125 mM); Tube 5, 8% polylysine with 2% agarose; and Tube 6, phosphate-buffered saline (PBS) buffer.

#### 2.1.3 Human participant cohorts

Eight healthy volunteers (age: 25 ± 5 years; 3 males, 5 females) were divided into two cohorts for different experimental purposes. The first cohort included four subjects whose data contributed to Studies 1-3. These participants were scanned at three CEST saturation power levels (0.5, 0.7, and 1.0 μT). The resulting multi-power dataset was used to demonstrate the image quality of whole-brain CEST imaging (Study 1), validate the conventional three-point $B_1$ correction method in vivo (Study 2), and to generate training and validation data for the $B_1$-correction neural network (Study 3). The second cohort consisted of another four healthy subjects who were each scanned twice on alternate days to assess the test-retest reproducibility of the proposed method (Study 4).

### 2.2 MRI data acquisition

All experiments were performed on a 3T whole-body MRI system (MAGNETOM Prisma Fit; Siemens Healthcare, Erlangen, Germany) using a 64-channel Head/Neck coil. Detailed data acquisition protocols for the phantom and human studies are provided below.

#### 2.2.1 Phantom scanning

CEST imaging of both phantoms was performed using a custom sequence comprising a pre-saturation CEST module, a fat-saturation module, and a single-shot 3D True FISP readout. The pre-saturation module consisted of 28 Gaussian-shaped radiofrequency (RF) pulses (100 ms each) with 5-ms intervals and crusher gradients between pulses to dephase residual transverse magnetization, yielding a total saturation time ($t_{Sat}$) of 3.0 seconds. To generate the Z-spectrum, this scheme was applied at 55 irradiation frequency offsets: two at –300 ppm for unsaturated reference images; ±50 ppm and ±35 ppm; from ±20 to ±11 ppm in 3-ppm steps; and from –10 to 10 ppm in 0.5-ppm steps. For each phantom, Z-spectra were acquired at three $B_1$ saturation amplitudes (0.5, 0.7, and 1.0 μT) to enable correction of $B_1$ field inhomogeneity via the three-point method.

The 3D True FISP readout used alternating-phase RF excitation with a preparatory half-flip-angle pulse to minimize signal oscillations. A short repetition time (TR) of 2 ms and echo time (TE) of 1 ms were chosen to enable rapid acquisition and reduce banding artifacts. At 3T, this TR shifts the first banding artifact to approximately ±2.0 ppm, an offset rarely observed in brain tissue. Data acquisition used centric spiral reordering, with the following parameters: flip angle = 25°, field of view (FOV) = 192 × 128 × 90 mm³, matrix size = 128 × 120 × 18, voxel size = 1.5 × 1.5 × 5.0 mm³.

Maps of the main magnetic field offset ($\Delta B_0$), relative transmit field ($rB_1$), and longitudinal relaxation time ($T_1$) were acquired in a 3.0-minute scan using a dual-echo modified four-angle method[21] (see Supporting Information, Table S1).

### 2.2.2 Human subject scanning

Both cohorts of human subjects were scanned using the True FISP–based CEST sequence, with pre-saturation parameters identical to those used in the phantom experiments. The True FISP readout was optimized for whole-brain coverage with sagittal orientation, TR = 2.0 ms, TE = 1.0 ms, flip angle = 25°, FOV = 220 × 220 × 200 mm³, matrix size = 88 × 88 × 80, and isotropic voxel size = 2.5 × 2.5 × 2.5 mm³. Acquisition was accelerated using generalized autocalibrating partially parallel acquisitions (GRAPPA) with a 2 × 2 acceleration factor and elliptical sampling, resulting in a readout time ($t_{RO}$) of 3.6 s. The total time per frequency offset was 6.6 s ($t_{Sat} + t_{RO}$), yielding an overall Z-spectrum scan time of approximately 6.0 minutes.

The first cohort (N=4) was scanned three times at different $B_1$ saturation amplitudes (0.5, 0.7, and 1.0 µT). In contrast, the second cohort (N=4) was scanned at a single $B_1$ amplitude of 0.7 µT, with the entire procedure repeated on two alternate days to assess test-retest reproducibility.

High-resolution whole-brain $T_1$-weighted images were acquired for both cohorts using a magnetization-prepared rapid gradient echo (MP-RAGE) sequence with TR = 2500 ms, TE = 2.2 ms, inversion time (TI) = 1000 ms, flip angle = 8°, FOV = 220 × 220 × 192 mm³, matrix size = 220 × 220 × 192, and isotropic voxel size = 1.0 × 1.0 × 1.0 mm³. Additionally, all human subjects underwent a rapid 3.0-minute scan using the dual-echo modified four-angle method to acquire $\Delta B_0$, $rB_1$, and $T_1$ maps.

For all human participants in both cohorts, the system continuously monitored the specific absorption rate (SAR) to ensure it remained well within the limits defined by the International Electrotechnical Commission (IEC) for normal operating mode.

### 2.3 Data processing and analysis

#### 2.3.1 Image pre-processing and correction map generation

All raw CEST images from human subjects underwent a multi-step pre-processing pipeline. First, to correct for head motion, all saturated images ($S_{sat}$) were rigidly aligned to the unsaturated reference image ($S_0$) using SPM 12 (v7771). Motion-corrected images were then denoised on a slice-by-slice basis using a deep-learning model developed by Chen H et al.[22].

In parallel, the images from the multi-echo, multi-angle acquisition of both phantoms and human subjects were used to generate the necessary correction maps. Specifically, the $\Delta B_0$ maps were derived from the dual-echo data, while the $rB_1$ and $T_1$ maps were calculated using the four-angle method as described by Bouhrara et al[21]. All correction maps ($\Delta B_0$, $rB_1$, $T_1$) were subsequently co-registered and resampled to match the spatial resolution and image space of the raw CEST images prior to further analysis.

#### 2.3.2 Z-spectrum correction

The pre-processed data were first normalized on a voxel-wise basis to generate the Z-spectrum for each voxel ($Z(\Delta\omega) = S_{sat}(\Delta\omega)/S_0$). Subsequently, the normalized Z-spectrum was corrected for $B_0$ field inhomogeneities by shifting each voxel's spectrum along the frequency axis according to its corresponding value in the $\Delta B_0$ map. Subsequently, transmit $B_1$ field correction was performed by using two distinct approaches depending on the study purpose. The conventional three-point correction method was applied for all phantom data to evaluate its correction performance (Study 2) and to the first human cohort to generate reference Z-spectra for the neural network training (Study 3). The trained neural network model, described in the following section, was applied to all data from the test-retest reproducibility cohort (Study 4).

#### 2.3.3 Neural network for rapid $B_1$ correction

To facilitate rapid $B_1$ correction in a clinical workflow, we developed and implemented a machine learning approach inspired by Hunger et al.[20]. A two-layer feed-forward neural network with 15 neurons in the hidden layer was implemented in MATLAB. The network was designed to emulate the results of the conventional three-point method using data from only a single $B_1$ power acquisition. For each voxel, the uncorrected Z-spectrum from the 0.7 μT scan and the corresponding $rB_1$ map value served as network inputs, while the Z-spectrum corrected by the three-point method served as the target output.

The model was trained and validated using data from the first cohort of four healthy subjects. Voxel-wise Z-spectra from three of these subjects were pooled and then randomly partitioned into a training set (70%), a validation set (15%), and a test set (15%). The validation set was used during the training process to monitor for overfitting. To robustly assess the final model's generalization ability, the data from the fourth subject was entirely held out as an independent test set for evaluation on unseen data.

**2.3.4 Quantitative analysis using Lorentzian fitting**

Fully corrected Z-spectra were analyzed voxel-wise using a four-pool Lorentzian model, assuming the Z-spectrum is a superposition of contributions from direct water saturation, MT, nuclear Overhauser enhancement (NOE), and amide proton transfer (APT). Following $B_0$ correction, the center of the direct water saturation pool was fixed at 0 ppm. The center frequencies of the other three pools were initialized at their expected offsets (-1.0 ppm for MT, -3.5 ppm for NOE, and +3.5 ppm for APT)[23]. To achieve optimal fitting, the center frequencies, amplitudes and widths of these pools, were treated as free parameters during the optimization process.

The model function was defined as:

$$Z(\Delta\omega) = 1 - \sum_{i=1}^{4} L_i(\Delta\omega), \quad \text{where} \quad L_i(\Delta\omega) = A_i \frac{W_i^2/4}{W_i^2/4 + (\Delta\omega - \Delta_i)^2}. \tag{1}$$

Here, $A_i$, $\Delta_i$, and $W_i$ are the amplitude, frequency offset, and full width at half maximum of the i-th Lorentzian pool, respectively. The initial values and fitting constraints for all model parameters are provided in Supporting Information (Table S2).

From the fitted model, two quantitative metrics were derived: the magnetization transfer ratio based on Lorentzian difference ($MTR_{LD}$) and the confounder-corrected AREX:

$$MTR_{LD}(\Delta\omega) = Z_{ref}(\Delta\omega) - Z_{lab}(\Delta\omega) \quad (2)$$

$$AREX(\Delta\omega)=(Z_{lab}(\Delta\omega)^{-1} - Z_{ref}(\Delta\omega)^{-1})/T_1 \quad (3)$$

Here, $T_1$ is the tissue longitudinal relaxation time, $Z_{lab}(\Delta\omega)$ is the total fitted Z-spectrum signal from the model, and $Z_{ref}(\Delta\omega)$ denotes the reference signal obtained by summing all Lorentzian pools except for the pool of interest (e.g., the APT pool).

### 2.3.5 ROI segmentation

Regional analysis was performed using an automated segmentation workflow. First, each subject's high-resolution $T_1$-weighted image was processed using the FreeSurfer analysis suite (v8.1.0) to perform automated volumetric segmentation and cortical parcellation. This process defined anatomical regions-of-interest (ROIs) based on the Desikan-Killiany atlas. After excluding non-brain tissue masks like cerebrospinal fluid, a total of 96 distinct cortical and subcortical regions were retained for analysis.

To align the CEST data with these anatomical ROIs, the unsaturated reference image ($S_0$) was co-registered to the $T_1$-weighted image to compute the spatial transformation matrix. This matrix was subsequently applied to all derived quantitative CEST maps (e.g., $MTR_{LD}$ and AREX maps), bringing them into the same anatomical space for regional quantification.

## 2.4 Statistical analysis

All statistical analyses were performed using custom scripts written in MATLAB (The MathWorks, Natick, MA).

For the phantom-based $B_1$ and $T_1$ correction validation (Study 2), the coefficient of variation (CV) was used to quantify data consistency. CVs of mean $MTR_{LD}$ values across the six identical tubes in phantom 1 were calculated to assess the three-point $B_1$ correction, while CVs of mean APT_AREX values across the five polylysine-containing tubes with varying $T_1$ and MT properties in phantom 2 were calculated to evaluate the AREX metric.

For the evaluation of the neural network model for $B_1$ correction (Study 3), training performance was assessed using the mean squared error (MSE) and the Pearson correlation coefficient (R) between the network output and the target data. To validate its accuracy in human subjects, the intraclass correlation coefficient (ICC) with a 95% confidence interval was calculated to determine the agreement between the regional $MTR_{LD}$ values obtained from the rapid neural network and those from the conventional three-point $B_1$ correction across all 96 brain regions.

For the assessment of test-retest reproducibility of whole-brain CEST metrics (Study 4), the CV for each of the 96 segmented brain regions was calculated from the mean CEST metric values within each region obtained from the two repeated scans of each subject. The reproducibility for each region was then reported as the average CV across the four subjects in the reproducibility cohort.

## 3 RESULTS

### 3.1 Image quality and safety in whole-brain CEST protocol

The proposed single-shot 3D True FISP sequence yielded high-quality whole-brain CEST images from all subjects. As shown in representative images from one subject (Figure 1), the unsaturated reference ($M_0$) images were characterized by a notable absence of visible banding artifacts, which can be attributed to the short repetition time (TR = 2 ms) of the sequence. In regions prone to magnetic susceptibility artifacts, such as near the nasal cavity, the $M_0$ images exhibited minimal geometric distortion, comparable to that observed in high-resolution $T_1$-weighted images acquired using the MP-RAGE sequence. The corresponding main magnetic field offset ($\Delta B_0$) map confirmed that the field shift was maintained below 2 ppm across the majority of brain tissue. For all human scans conducted at a $B_1$ amplitude of 0.7 μT, the average specific absorption rate (SAR) was 52% ± 6.8% of the IEC normal operating mode limit, ensuring the protocol operated within prescribed safety guidelines.

### 3.2 Phantom-based validation of correction methods

#### 3.2.1 $B_1$ field inhomogeneity correction

The effectiveness of the three-point $B_1$ correction method was demonstrated in a homogeneous phantom, with the results presented in Figure 2. Prior to correction, the magnetization transfer ($MT_MTR_{LD}$) map (Figure 2C) displayed a spatial variation that corresponded with the pattern of the measured $rB_1$ inhomogeneity (Figure 2D). Following the application of the three-point $B_1$ correction, this spatial dependency was reduced, yielding more uniform $MT_MTR_{LD}$ values across the six identical test tubes (Figure 2F).

This effect was quantified using the CV across the tubes, as detailed in Table 1. The CV for $MT_MTR_{LD}$ decreased from 22.49% before correction to 4.61% after correction. In contrast, the CV for $APT_MTR_{LD}$ was 4.44% before correction and remained at a similar level (5.16%) after the procedure.

### 3.2.2 Correction for $T_1$ and MT confounders with the AREX metric

The AREX metric effectively corrected for signal variability caused by $T_1$ and MT confounders in the custom-designed phantom. After the three-point $B_1$ correction was applied, the $APT_MTR_{LD}$ map (Figure 3B) exhibited notable signal variations among the tubes with varying $T_1$ and MT properties. This variability is further illustrated in the corresponding signal distribution histogram (Figure 3C), which shows five distinct distributions for the polylysine tubes. In contrast, the APT_AREX metric, which incorporates the measured $T_1$ map (Figure 3D), yielded consistent signal intensities across all five polylysine tubes (Figure 3E). The histogram in Figure 3F confirms this observation, showing that the five distributions became tightly clustered. Quantitatively, the CV across the five polylysine-containing tubes was reduced from 33.6% for $APT_MTR_{LD}$ to 6.9% for APT_AREX, indicating that the metric effectively corrected for the confounding factors.

## 3.3 In vivo validation of field and confounder corrections

The effects of $B_1$ inhomogeneity and the performance of the correction pipeline in the human brain are demonstrated in Figure 4, using data from a representative healthy subject in the first cohort. Without $B_1$ correction, the $MTR_{LD}$ maps exhibited spatial artifacts that corresponded to the underlying $rB_1$ inhomogeneity (Figure 4, top and bottom rows). For example, pronounced signal loss was observed in the frontal lobe of the uncorrected axial APT and MT maps, a region coinciding with lower transmit $B_1$ fields (Figure 4, red arrows).

Application of the three-point $B_1$ correction method substantially reduced these $B_1$-dependent artifacts, resulting in more homogeneous quantitative maps (Figure 4, second row). The magnitude of the applied correction is visualized in the difference maps (Figure 4, third row).

Further correction for biological confounders using the AREX metric resulted in maps with enhanced gray-white matter contrast. As shown in the resulting AREX maps (Figure 4, fourth row), this enhancement was particularly evident for NOE and MT when compared to the $B_1$-corrected $MTR_{LD}$ maps.

### 3.4 Validation of the neural network for rapid $B_1$ correction

#### 3.4.1 Network training and performance

The performance of the trained model is summarized in Table 2. After 18 hours of training, the correlation coefficient (R) between the network output and the target data exceeded 0.9997 for the training, validation, and testing datasets. The MSE was approximately 1.7-1.8 × $10^{-5}$ across all three datasets.

#### 3.4.2 In vivo comparison with the three-point $B_1$ correction method

The performance of the trained neural network was validated in a healthy subject whose data were excluded from the training process. A strong visual agreement was observed between the $MTR_{LD}$ maps corrected by the neural network and those corrected by the conventional three-point method, which served as the reference standard (Figure 5). The difference images, calculated by subtracting the reference results from the neural network results, showed minimal residual values for APT, NOE, and MT effects, further indicating a high degree of similarity between the two approaches.

This agreement was further quantified through a regional analysis across 96 brain regions. The intraclass correlation coefficient (ICC) indicated strong to excellent agreement for all effects: $APT_MTR_{LD}$ (ICC = 0.975 [95% CI: 0.963, 0.983]), $NOE_MTR_{LD}$ (ICC = 0.985 [95% CI: 0.977, 0.990]), and $MT_MTR_{LD}$ (ICC = 0.997 [95% CI: 0.995, 0.998]). As shown in Figure 6, scatter plots of the regional $MTR_{LD}$ values confirmed a strong linear correlation for all three CEST effects (R > 0.97). Furthermore, Bland-Altman analysis revealed a mean difference (bias) centered near zero with narrow limits of agreement, confirming the absence of systematic bias between the neural network method and the reference three-point correction.

### 3.5 Final quantitative maps and pipeline reproducibility

#### 3.5.1 Representative quantitative maps

Figure 7 displays the final multi-parameter quantitative maps from a representative healthy subject from the second cohort. These maps represent the output of the complete proposed pipeline, which includes corrections for both $B_0$ and $B_1$ field inhomogeneities, with the $B_1$ correction performed using the validated neural network model. The figure presents a side-by-side comparison of the quantitative maps derived from (A) the $MTR_{LD}$ metric and (B) the AREX metric.

#### 3.5.2 Test-retest reproducibility

The final proposed pipeline demonstrated high test-retest reproducibility. The analysis of the APT_AREX metric, for instance, showed excellent stability across 96 segmented brain regions (Figure 8). The average coefficient of variation (CV) for APT_AREX was less than 10% for 92 out of the 96 (95.8%) analyzed regions. High stability was also observed for other quantitative metrics, including NOE_AREX, MT_AREX, and all $MTR_{LD}$ variants (Supporting Information, Figures S1-S5).

## 4 DISCUSSION

In this study, we developed and validated a comprehensive solution for whole-brain, multi-pool CEST imaging at 3T that is rapid, accurate, and highly reproducible. Our approach successfully integrates a single-shot 3D True FISP readout sequence with a streamlined workflow for $B_0$, $B_1$, and $T_1$ correction, highlighted by the implementation of a neural network for rapid $B_1$ field inhomogeneity correction. By addressing several critical bottlenecks that have historically hindered the clinical translation of CEST, our work provides a practical pathway for its application in diseases affecting multiple brain regions, such as neurodegenerative disorders.

A cornerstone of our method is the use of a single-shot 3D True FISP readout. While our previous work had already demonstrated the superior SNR of this sequence for limited multi-slice acquisitions[19], the present study successfully optimized and extended this approach for true whole-brain coverage. This high SNR afforded us the flexibility to significantly shorten the acquisition time by using a very short TR and a high parallel imaging acceleration factor. This efficiency resulted in a total acquisition time of approximately 6.0 minutes for a 55-point Z-spectrum acquired with the True FISP sequence. In addition, the necessary correction maps were acquired in just 3.0 minutes using a dedicated rapid mapping protocol, bringing the total scan time to a clinically feasible 9 minutes. Furthermore, the short TR also effectively mitigated the banding artifacts typically associated with True FISP techniques[24]. Specifically, reducing the TR to 2 ms shifted the off-resonance frequency at which banding occurs to approximately ±2.0 ppm. This threshold is far beyond the typical magnetic field variations observed across the vast majority of brain tissue (Figure 1), enabling us to acquire high-quality, whole-brain images even in regions prone to magnetic susceptibility artifacts.

The accuracy of CEST imaging, however, is critically dependent on correcting for $B_1$ transmit field inhomogeneities. Traditional correction methods, which require acquiring the Z-spectrum at multiple power levels, are prohibitively time-consuming[16]. Our study tackled this challenge systematically. First, we confirmed the efficacy of the established three-point $B_1$ correction method in phantoms. The correction proved highly effective at removing $B_1$-dependent variations, as demonstrated by the CV for $MT_MTR_{LD}$ decreasing significantly from 22.49% to 4.61% post-correction. This observation highlights the critical need for $B_1$ correction, especially for highly $B_1$-sensitive metrics like the MT effect, and provided a reliable reference for subsequent comparisons. Interestingly, we also observed that the impact of the correction varied between different CEST pools. The CV for $APT_MTR_{LD}$, which was already low, showed a slight increase after correction (from 4.44% to 5.16%). This is likely because the APT effect at the applied $B_1$ power is less sensitive to $B_1$ inhomogeneity than the MT effect; thus, the marginal benefit from correction was potentially outweighed by minor noise propagation inherent to the algorithm itself. This finding does not diminish the need for $B_1$ correction but underscores its nuanced behavior and importance for certain quantitative metrics. Building on this, we successfully implemented and validated a custom-trained, two-layer feed-forward neural network that accurately replicates the results of the three-point method in human subjects (ICC > 0.97), but without the need for additional scans at different power levels. This integration effectively eliminates the prolonged scan times associated with multi-power acquisitions, making comprehensive $B_1$ correction feasible within a clinical workflow.

Beyond field corrections, the interpretability of CEST metrics is often confounded by factors like $T_1$ relaxation and MT effects. Our work validates the utility of the AREX metric to mitigate these confounders. In a phantom designed with varying $T_1$ and MT properties, the APT_$MTR_{LD}$ metric showed significant variability (CV = 33.6%) that did not reflect the true chemical exchange. In contrast, the APT_AREX metric successfully corrected for these effects, yielding consistent values across the tubes (CV = 6.9%) and isolating the desired exchange effect. In human subjects, this translated to AREX maps consistently showing enhanced gray-white matter contrast compared to $MTR_{LD}$ maps, suggesting improved biological specificity.

Finally, we rigorously assessed test-retest reproducibility, establishing a benchmark for the field. Our results demonstrated high reproducibility at the brain-region level for all derived CEST metrics, including both AREX and $MTR_{LD}$ variants (Figure 8 and Supplementary Figures S1-S5). Across all metrics, the majority of the 96 brain regions showed excellent stability, with average CVs largely falling below 10%. However, a consistent pattern of higher variability was observed in a specific set of anatomical regions across these different metrics. These typically included the temporal poles, entorhinal cortices, parts of the corpus callosum, and other cortical or subcortical structures adjacent to large CSF spaces (e.g., caudate, cuneus). The increased variability in these regions likely stems from their susceptibility to a confluence of known neuroimaging challenges. For instance, the high variability consistently seen in the temporal poles and entorhinal cortices is largely driven by heightened magnetic susceptibility artifacts near air-tissue interfaces like the sinuses. In addition, a more general factor affecting many of the other identified regions—including the caudate, corpus callosum, and cuneus—is their direct proximity to large CSF spaces. Pulsatile CSF flow and partial volume effects at these tissue-fluid interfaces are known to introduce significant signal instability, posing a common challenge for quantitative MRI[25,26]. Despite these localized challenges, the high level of robustness demonstrated across the vast majority of the brain is essential for clinical applications, particularly for longitudinal monitoring of disease progression or treatment response, and for cross-sectional studies comparing different patient groups[27,28].

This study has several limitations. First, the validation was conducted exclusively in healthy volunteers, and the sample size for our human cohorts was small (N=4 for each group). This is a primary limitation of the current study. Consequently, while our results demonstrate high reproducibility and accuracy, the method's clinical utility and the neural network model's generalizability must be further validated in future studies involving larger and more diverse cohorts, including various patient populations, to confirm its overall stability and robustness. Second, while the total acquisition time of approximately 9 minutes is clinically acceptable, further acceleration could improve workflow efficiency. The analysis also relied on a simplified four-pool Lorentzian model; more advanced modeling could enhance the specificity of the derived metrics. Finally, the choice of a 3-second saturation time represents a trade-off between achieving a saturation steady-state and maintaining a clinically feasible scan time. This duration may not achieve a full steady state, which could affect its absolute quantitative accuracy as the $T_1$ correction within AREX relies on a steady-state assumption. Nevertheless, this trade-off was necessary for the rapid protocol targeted in this study. Future work could explore overcoming this limitation without extending the scan time by employing advanced reconstruction techniques like Quasi-steady-state (QUASS) reconstruction[3].

## 5 CONCLUSIONS

In conclusion, this study presents a validated, end-to-end solution for whole-brain multi-pool CEST imaging. By combining an efficient True FISP sequence with a rapid, neural network-driven correction pipeline and robust quantitative analysis, our method overcomes major barriers to clinical adoption. It delivers high-fidelity, reproducible, and quantitative multi-parameter maps of brain metabolism in a clinically acceptable timeframe, holding significant promise for advancing neuro-imaging research and diagnostics.

## ACKNOWLEDGMENTS

This study was supported by the Science and Technology Commission of Shanghai Municipality (Grant No. YDZX20243100003001001) and the STI 2030—Major Projects (No. 2021ZD0200500).

## DATA AVAILABILITY STATEMENT

The data that support the findings of this study are available from the corresponding author upon reasonable request. The custom CEST sequence used in this research can be made available to other institutions through the Siemens C2P (Customer to Partner) Exchange platform.

TABLE 1 Coefficient of variation (CV) of CEST metrics before and after three-point $B_1$ correction in a phantom.

| Metric | CV Before $B_1$ Correction | CV After $B_1$ Correction |
|---|---|---|
| $APT_MTR_{LD}$ | 4.44% | 5.16% |
| $MT_MTR_{LD}$ | 22.49% | 4.61% |

TABLE 2 Performance metrics of the neural network model for $B_1$ correction.

| Dataset | Sample size | MSE | R |
|---|---|---|---|
| Training | 215350 | $1.7540 \times 10^{-5}$ | 0.99975 |
| Validation | 46147 | $1.7475 \times 10^{-5}$ | 0.99975 |
| Testing | 46147 | $1.7932 \times 10^{-5}$ | 0.99974 |

Abbreviations: MSE, mean squared error; R, correlation coefficient.

**Figure Captions**

FIGURE 1 Image quality assessment of the whole-brain True FISP acquisition. The figure displays representative images from a healthy participant. The top row shows unsaturated reference images ($M_0$) acquired with the single-shot 3D True FISP sequence, which exhibit high quality with minimal banding artifacts and susceptibility-induced distortion. The middle row contains corresponding high-resolution $T_1$-weighted ($T_1W$) images for anatomical reference. The bottom row presents the main magnetic field offset ($\Delta B_0$) maps. $M_0$, unsaturated reference image; True FISP, true fast imaging with steady-state precession; $T_1W$, $T_1$-weighted; $\Delta B_0$, main magnetic field offset.

FIGURE 2 Validation of the three-point $B_1$ correction method in a homogeneous phantom. (A) Reference image of the phantom, which consists of six identical tubes (8% polylysine, 1% agarose, pH=7.3). (B, C) The amide proton transfer ($APT_MTR_{LD}$) and magnetization transfer ($MT_MTR_{LD}$) maps before $B_1$ correction. (D) The measured relative transmit field ($rB_1$) map, which shows significant inhomogeneity. (E, F) The corresponding $APT_MTR_{LD}$ and $MT_MTR_{LD}$ maps after applying the three-point $B_1$ correction. The correction visibly homogenizes the MT map, removing the spatial dependency on the $rB_1$ distribution. APT, amide proton transfer; $MTR_{LD}$, magnetization transfer ratio based on Lorentzian difference; MT, magnetization transfer; $rB_1$, relative transmit field.

FIGURE 3 Validation of the AREX metric in correcting for $T_1$- and MT-induced signal variability in a multi-component phantom. (A) Reference MR image of the phantom with labeled tube compositions. Tubes 1-5 contain 8% polylysine with varying $T_1$ and MT properties, while tube 6 contains PBS. (B) The $APT_MTR_{LD}$ map shows significant signal variability across tubes 1-5, reflected in the widely spread distributions in the corresponding histogram (C). (D) The measured $T_1$ map confirms $T_1$ variation across the tubes. (E) In contrast, the APT_AREX map demonstrates consistent signal intensity across tubes 1-5, which is confirmed by the tightly clustered distributions in its histogram (F). APT, amide proton transfer; $MTR_{LD}$, magnetization transfer ratio based on Lorentzian difference; AREX, apparent exchange-dependent relaxation; MT, magnetization transfer; PL, polylysine; Gd, gadolinium contrast agent; AG, agarose; PBS, phosphate-buffered saline.

FIGURE 4 In vivo $B_1$ correction and AREX quantification in a healthy subject. The figure displays whole-brain quantitative maps for amide proton transfer (APT), nuclear Overhauser enhancement (NOE), and magnetization transfer (MT) from a representative subject. The first row shows $MTR_{LD}$ maps without $B_1$ correction (Uncorrected), which exhibit conspicuous artifacts corresponding to the $rB_1$ inhomogeneity shown in the bottom row. The second row displays the $MTR_{LD}$ maps after applying the three-point $B_1$ correction. The third row visualizes the voxel-wise difference (Corrected - Uncorrected), highlighting the magnitude of the correction. The fourth row presents the final AREX maps, which are corrected for $T_1$ and MT confounders. The bottom row provides the reference anatomical ($T_1W$ image), relative transmit field ($rB_1$ map), and $T_1$ maps. $MTR_{LD}$, magnetization transfer ratio based on Lorentzian difference; AREX, apparent exchange-dependent relaxation; APT, amide proton transfer; NOE, nuclear Overhauser enhancement; MT, magnetization transfer; $rB_1$, relative transmit field.

FIGURE 5 Comparison of $B_1$ correction using the neural network versus the three-point method in a healthy subject. The figure displays whole-brain $MTR_{LD}$ maps for APT, NOE, and MT effects. The top row shows the $MTR_{LD}$ maps generated using the 3-point method, which serves as the reference. The middle row shows the corresponding maps generated using the Neural Network (NN) method. The bottom row presents the voxel-wise difference images (NN - 3-point), which show minimal residual differences between the two approaches. $MTR_{LD}$, magnetization transfer ratio based on Lorentzian difference; NN, neural network; APT, amide proton transfer; NOE, nuclear Overhauser enhancement; MT, magnetization transfer.

FIGURE 6 Quantitative comparison of the neural network and three-point $B_1$ correction methods in a healthy subject. The top row displays scatter plots of regional $MTR_{LD}$ values for amide proton transfer (APT), nuclear Overhauser enhancement (NOE), and magnetization transfer (MT) across 96 brain regions, comparing the neural network (NN) method with the three-point method. The solid black lines represent the linear fits, and the dashed green lines represent the line of identity. The bottom row presents the corresponding Bland-Altman plots, showing the difference between the two methods against their average. The dashed red line indicates the mean difference (bias), and the dotted black lines indicate the 95% limits of agreement (±1.96 SD). NN, neural network; $MTR_{LD}$, magnetization transfer ratio based on Lorentzian difference; APT, amide proton transfer; NOE, nuclear Overhauser enhancement; MT, magnetization transfer; SD, standard deviation.

FIGURE 7 Comparison of final quantitative maps derived from $MTR_{LD}$ and AREX metrics of the second cohort. Representative whole-brain maps for amide proton transfer (APT), nuclear Overhauser enhancement (NOE), and magnetization transfer (MT) from a healthy subject are shown. (A) Quantitative maps calculated using the $MTR_{LD}$ metric. (B) Corresponding maps calculated using the AREX metric. All maps were generated using the complete processing pipeline, including neural network-based $B_1$ correction. $MTR_{LD}$, magnetization transfer ratio based on Lorentzian difference; AREX, apparent exchange-dependent relaxation; APT, amide proton transfer; NOE, nuclear Overhauser enhancement; MT, magnetization transfer.

FIGURE 8 Test-retest reproducibility of the APT_AREX metric across human brain regions. The bar chart displays the mean coefficient of variation (CV) across 96 segmented brain regions, sorted in ascending order. Data were acquired from four healthy subjects, each scanned twice on alternate days. Each bar represents the mean CV across the four subjects, and the error bars indicate the standard deviation (SD). The analysis used the complete pipeline with neural network-based $B_1$ correction. Brain region abbreviations correspond to the FreeSurfer segmentation output (e.g., CC, Corpus Callosum). APT, amide proton transfer; AREX, apparent exchange-dependent relaxation; CV, coefficient of variation; SD, standard deviation.

**Supporting Information Captions**

TABLE S1 MRI acquisition parameters for $B_0$, $B_1$, and $T_1$ mapping.

TABLE S2 Initial values and boundaries for the center frequency (Δ), amplitude (A), and full width at half maximum (W) of the pools in the Lorentzian fitting model.

FIGURE S1 Test-retest reproducibility of the NOE_AREX metric across human brain regions. The bar chart displays the mean coefficient of variation (CV) across 96 segmented brain regions, sorted in ascending order. Data were acquired from four healthy subjects, each scanned twice on alternate days. Each bar represents the mean CV across the four subjects, and the error bars indicate the standard deviation (SD). The analysis used the complete pipeline with neural network-based $B_1$ correction. Brain region abbreviations correspond to the FreeSurfer segmentation output (e.g., CC, Corpus Callosum). NOE, nuclear Overhauser enhancement; AREX, apparent exchange-dependent relaxation; CV, coefficient of variation; SD, standard deviation.

FIGURE S2 Test-retest reproducibility of the MT_AREX metric across human brain regions. The bar chart displays the mean coefficient of variation (CV) across 96 segmented brain regions, sorted in ascending order. Data were acquired from four healthy subjects, each scanned twice on alternate days. Each bar represents the mean CV across the four subjects, and the error bars

indicate the standard deviation (SD). The analysis used the complete pipeline with neural network-based $B_1$ correction. Brain region abbreviations correspond to the FreeSurfer segmentation output (e.g., CC, Corpus Callosum). MT, magnetization transfer; AREX, apparent exchange-dependent relaxation; CV, coefficient of variation; SD, standard deviation.

FIGURE S3 Test-retest reproducibility of the APT_$MTR_{LD}$ metric across human brain regions. The bar chart displays the mean coefficient of variation (CV) across 96 segmented brain regions, sorted in ascending order. Data were acquired from four healthy subjects, each scanned twice on alternate days. Each bar represents the mean CV across the four subjects, and the error bars indicate the standard deviation (SD). The analysis used the complete pipeline with neural network-based $B_1$ correction. Brain region abbreviations correspond to the FreeSurfer segmentation output (e.g., CC, Corpus Callosum). APT, amide proton transfer; $MTR_{LD}$, magnetization transfer ratio based on Lorentzian difference; CV, coefficient of variation; SD, standard deviation.

FIGURE S4 Test-retest reproducibility of the NOE_$MTR_{LD}$ metric across human brain regions. The bar chart displays the mean coefficient of variation (CV) across 96 segmented brain regions, sorted in ascending order. Data were acquired from four healthy subjects, each scanned twice on alternate days. Each bar represents the mean CV across the four subjects, and the error bars indicate the standard deviation (SD). The analysis used the complete pipeline with neural network-based $B_1$ correction. Brain region abbreviations correspond to the FreeSurfer segmentation output (e.g., CC, Corpus Callosum). NOE, nuclear Overhauser enhancement; $MTR_{LD}$, magnetization transfer ratio based on Lorentzian difference; CV, coefficient of variation; SD, standard deviation.

FIGURE S5 Test-retest reproducibility of the MT_$MTR_{LD}$ metric across human brain regions. The bar chart displays the mean coefficient of variation (CV) across 96 segmented brain regions, sorted in ascending order. Data were acquired from four healthy subjects, each scanned twice on alternate days. Each bar represents the mean CV across the four subjects, and the error bars

indicate the standard deviation (SD). The analysis used the complete pipeline with neural network-based $B_1$ correction. Brain region abbreviations correspond to the FreeSurfer segmentation output (e.g., CC, Corpus Callosum). MT, magnetization transfer; $MTR_{LD}$, magnetization transfer ratio based on Lorentzian difference; CV, coefficient of variation; SD, standard deviation.

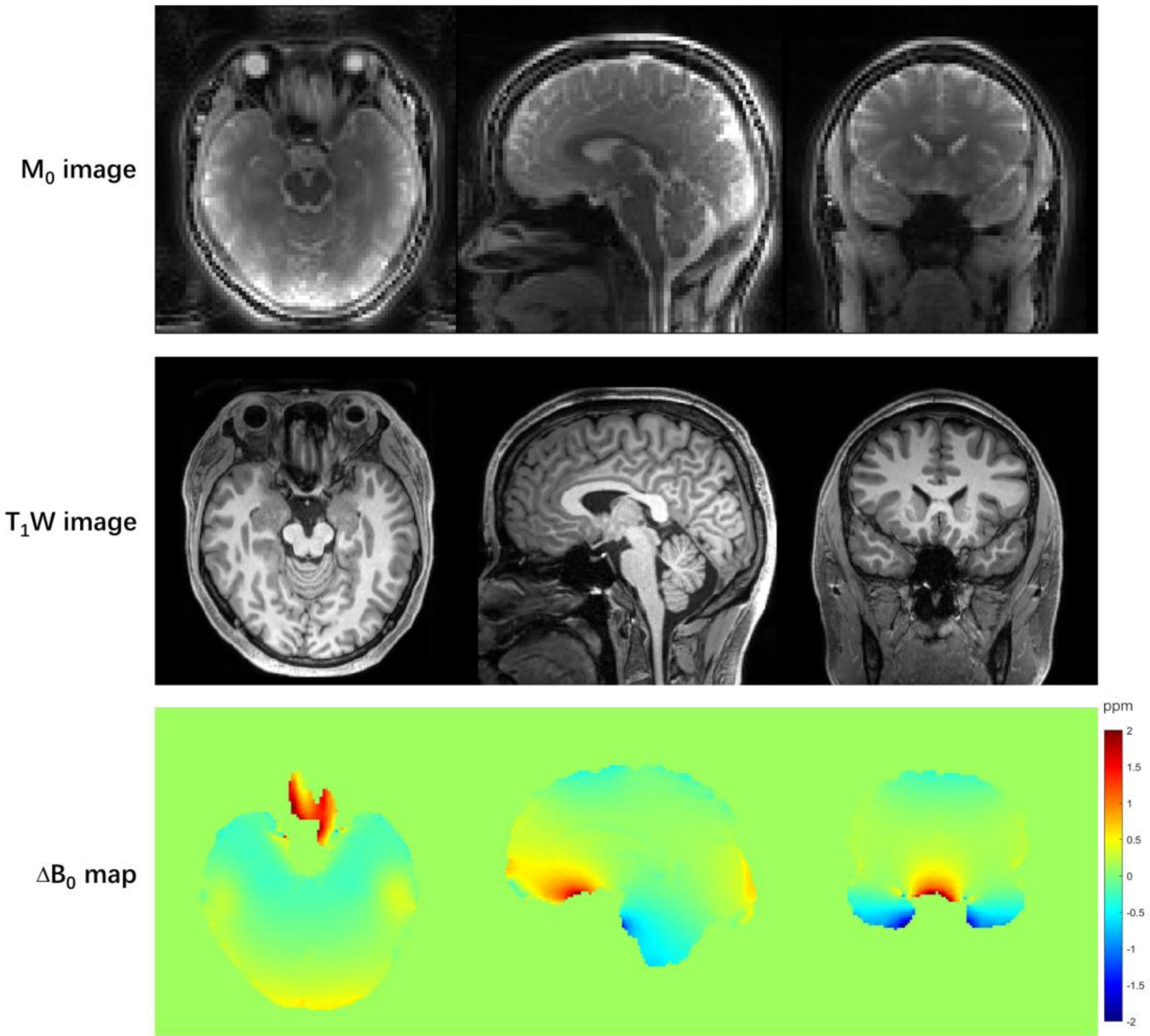

FIGURE 1 Image quality assessment of the whole-brain True FISP acquisition. The figure displays representative images from a healthy participant. The top row shows unsaturated reference images ($M_0$) acquired with the single-shot 3D True FISP sequence, which exhibit high quality with minimal banding artifacts and susceptibility-induced distortion. The middle row contains corresponding high-resolution $T_1$-weighted ($T_1W$) images for anatomical reference. The bottom row presents the main magnetic field offset ($\Delta B_0$) maps. $M_0$, unsaturated reference image; True FISP, true fast imaging with steady-state precession; $T_1W$, $T_1$-weighted; $\Delta B_0$, main magnetic field offset.

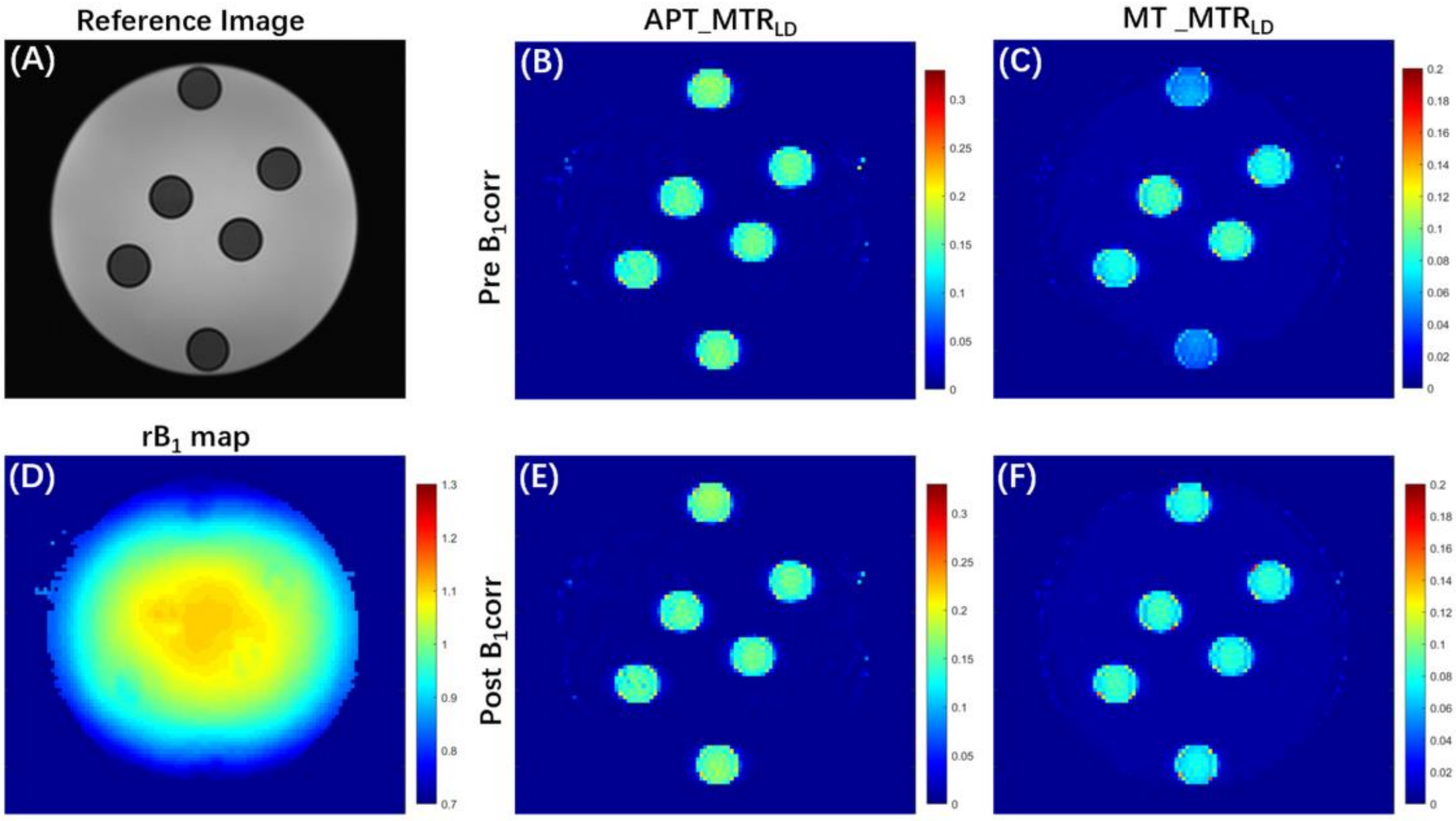

FIGURE 2 Validation of the three-point $B_1$ correction method in a homogeneous phantom. (A) Reference image of the phantom, which consists of six identical tubes (8% polylysine, 1% agarose, pH=7.3). (B, C) The amide proton transfer (APT_MTR$_{LD}$) and magnetization transfer (MT_MTR$_{LD}$) maps before $B_1$ correction. (D) The measured relative transmit field (r$B_1$) map, which shows significant inhomogeneity. (E, F) The corresponding APT_MTR$_{LD}$ and MT_MTR$_{LD}$ maps after applying the three-point $B_1$ correction. The correction visibly homogenizes the MT map, removing the spatial dependency on the r$B_1$ distribution. APT, amide proton transfer; MTR$_{LD}$, magnetization transfer ratio based on Lorentzian difference; MT, magnetization transfer; r$B_1$, relative transmit field.

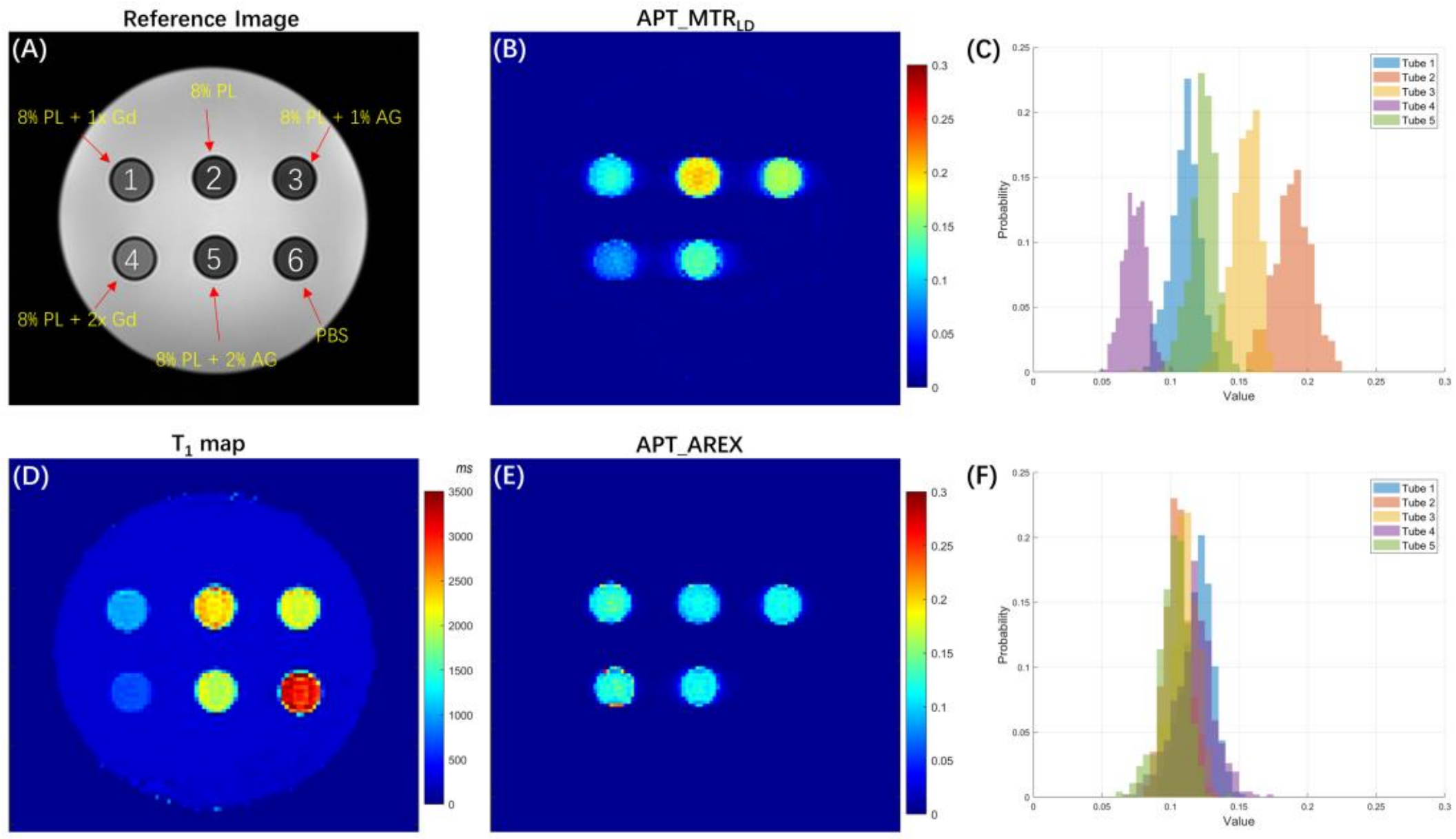

FIGURE 3 Validation of the AREX metric in correcting for $T_1$- and MT-induced signal variability in a multi-component phantom. (A) Reference MR image of the phantom with labeled tube compositions. Tubes 1-5 contain 8% polylysine with varying $T_1$ and MT properties, while tube 6 contains PBS. (B) The APT_$MTR_{LD}$ map shows significant signal variability across tubes 1-5, reflected in the widely spread distributions in the corresponding histogram (C). (D) The measured $T_1$ map confirms $T_1$ variation across the tubes. (E) the APT_AREX map demonstrates consistent signal intensity across tubes 1-5, which is confirmed by the tightly clustered distributions in its histogram (F). APT, amide proton transfer; $MTR_{LD}$, magnetization transfer ratio based on Lorentzian difference; AREX, apparent exchange-dependent relaxation; MT, magnetization transfer; PL, polylysine; Gd, gadolinium contrast agent; AG, agarose; PBS, phosphate-buffered saline.

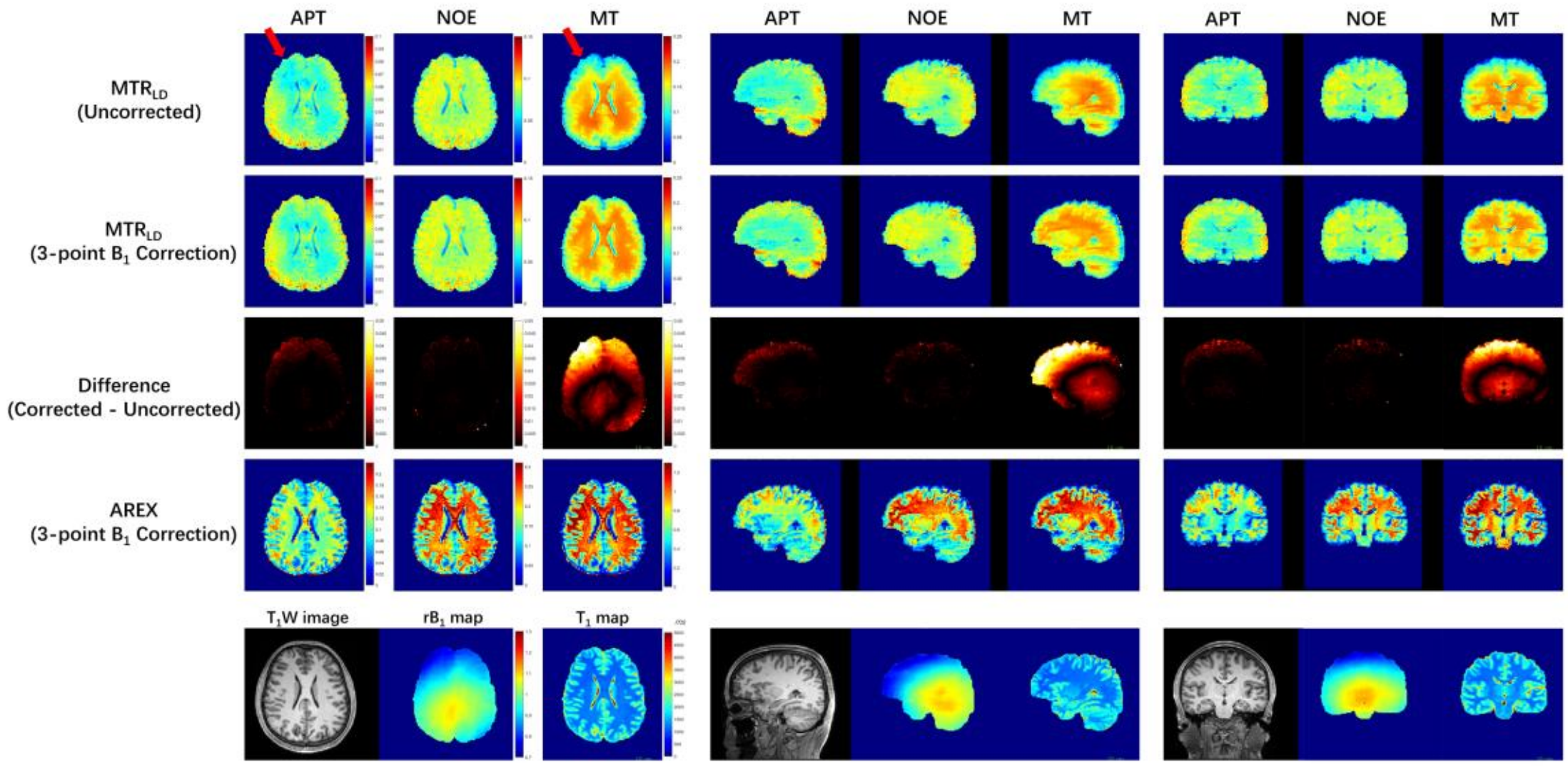

FIGURE 4 In vivo $B_1$ correction and AREX quantification in a healthy subject. The figure displays whole-brain quantitative maps for amide proton transfer (APT), nuclear Overhauser enhancement (NOE), and magnetization transfer (MT) from a representative subject. The first row shows $MTR_{LD}$ maps without $B_1$ correction (Uncorrected), which exhibit conspicuous artifacts corresponding to the $rB_1$ inhomogeneity shown in the bottom row. The second row displays the $MTR_{LD}$ maps after applying the three-point $B_1$ correction. The third row visualizes the voxel-wise difference (Corrected - Uncorrected), highlighting the magnitude of the correction. The fourth row presents the final AREX maps, which are corrected for $T_1$ and MT confounders. The bottom row provides the reference anatomical ($T_1$W image), relative transmit field ($rB_1$ map), and $T_1$ maps. $MTR_{LD}$, magnetization transfer ratio based on Lorentzian difference; AREX, apparent exchange-dependent relaxation; APT, amide proton transfer; NOE, nuclear Overhauser enhancement; MT, magnetization transfer; $rB_1$, relative transmit field.

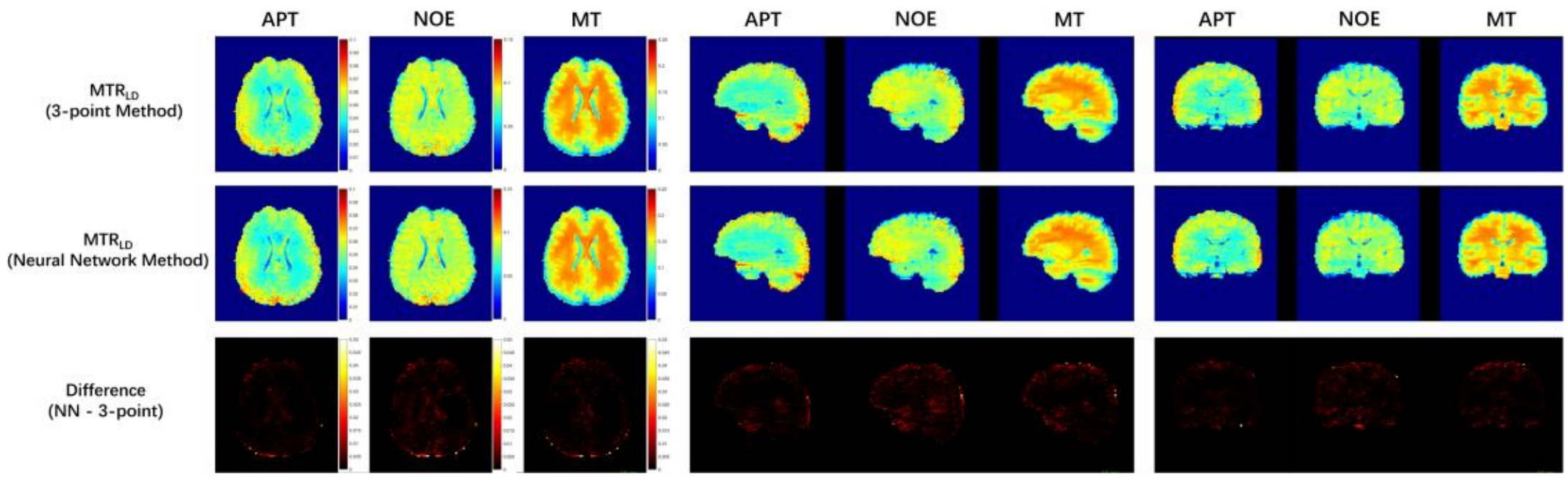

FIGURE 5 Comparison of $B_1$ correction using the neural network versus the three-point method in a healthy subject. The figure displays whole-brain $MTR_{LD}$ maps for APT, NOE, and MT effects. The top row shows the $MTR_{LD}$ maps generated using the 3-point method, which serves as the reference. The middle row shows the corresponding maps generated using the Neural Network (NN) method. The bottom row presents the voxel-wise difference images (NN - 3-point), which show minimal residual differences between the two approaches. $MTR_{LD}$, magnetization transfer ratio based on Lorentzian difference; NN, neural network; APT, amide proton transfer; NOE, nuclear Overhauser enhancement; MT, magnetization transfer.

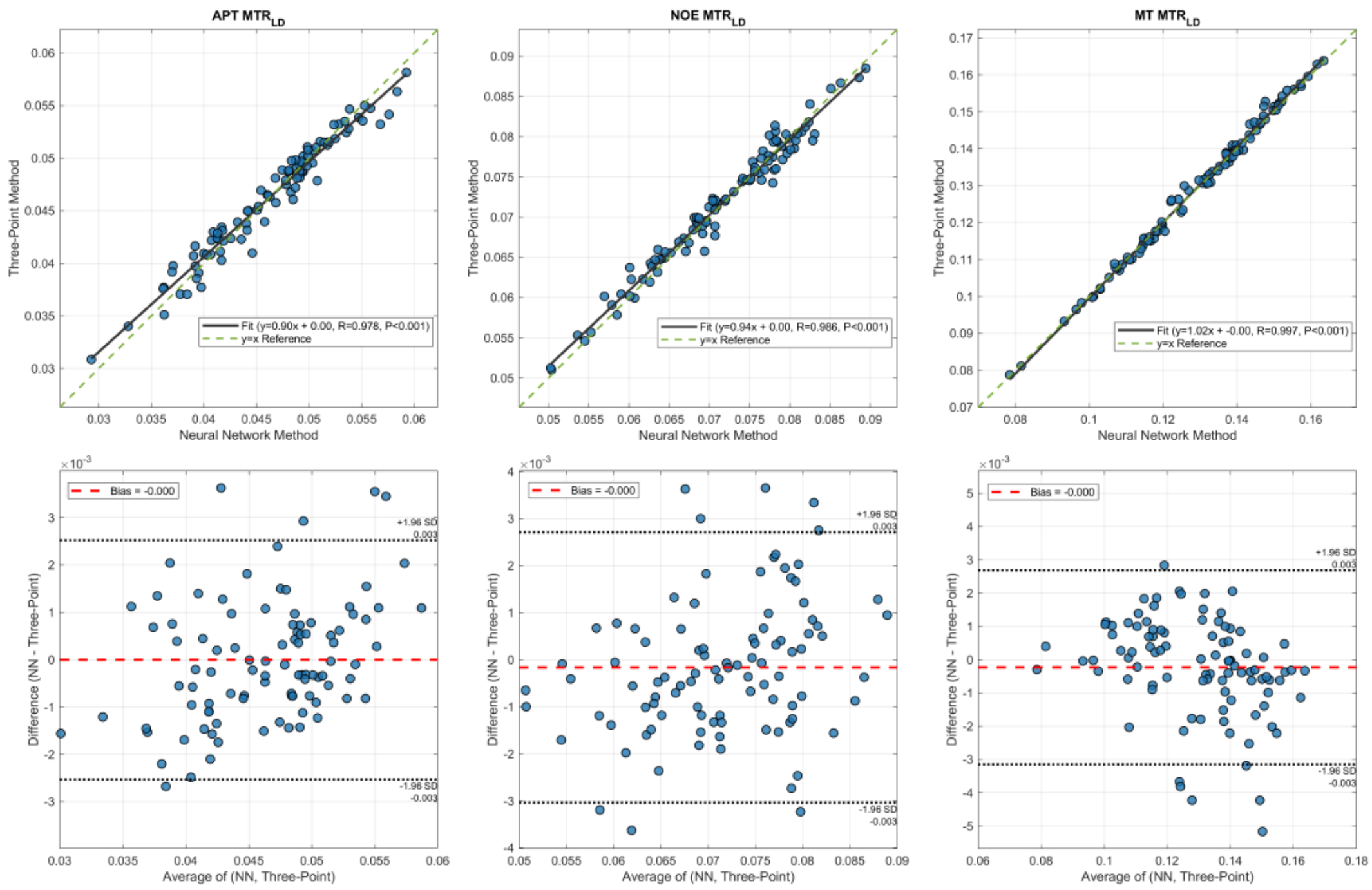

FIGURE 6 Quantitative comparison of the neural network and three-point $B_1$ correction methods in a healthy subject. The top row displays scatter plots of regional $MTR_{LD}$ values for amide proton transfer (APT), nuclear Overhauser enhancement (NOE), and magnetization transfer (MT) across 96 brain regions, comparing the neural network (NN) method with the three-point method. The solid black lines represent the linear fits, and the dashed green lines represent the line of identity. The bottom row presents the corresponding Bland-Altman plots, showing the difference between the two methods against their average. The dashed red line indicates the mean difference (bias), and the dotted black lines indicate the 95% limits of agreement (±1.96 SD). NN, neural network; $MTR_{LD}$, magnetization transfer ratio based on Lorentzian difference; APT, amide proton transfer; NOE, nuclear Overhauser enhancement; MT, magnetization transfer; SD, standard deviation.

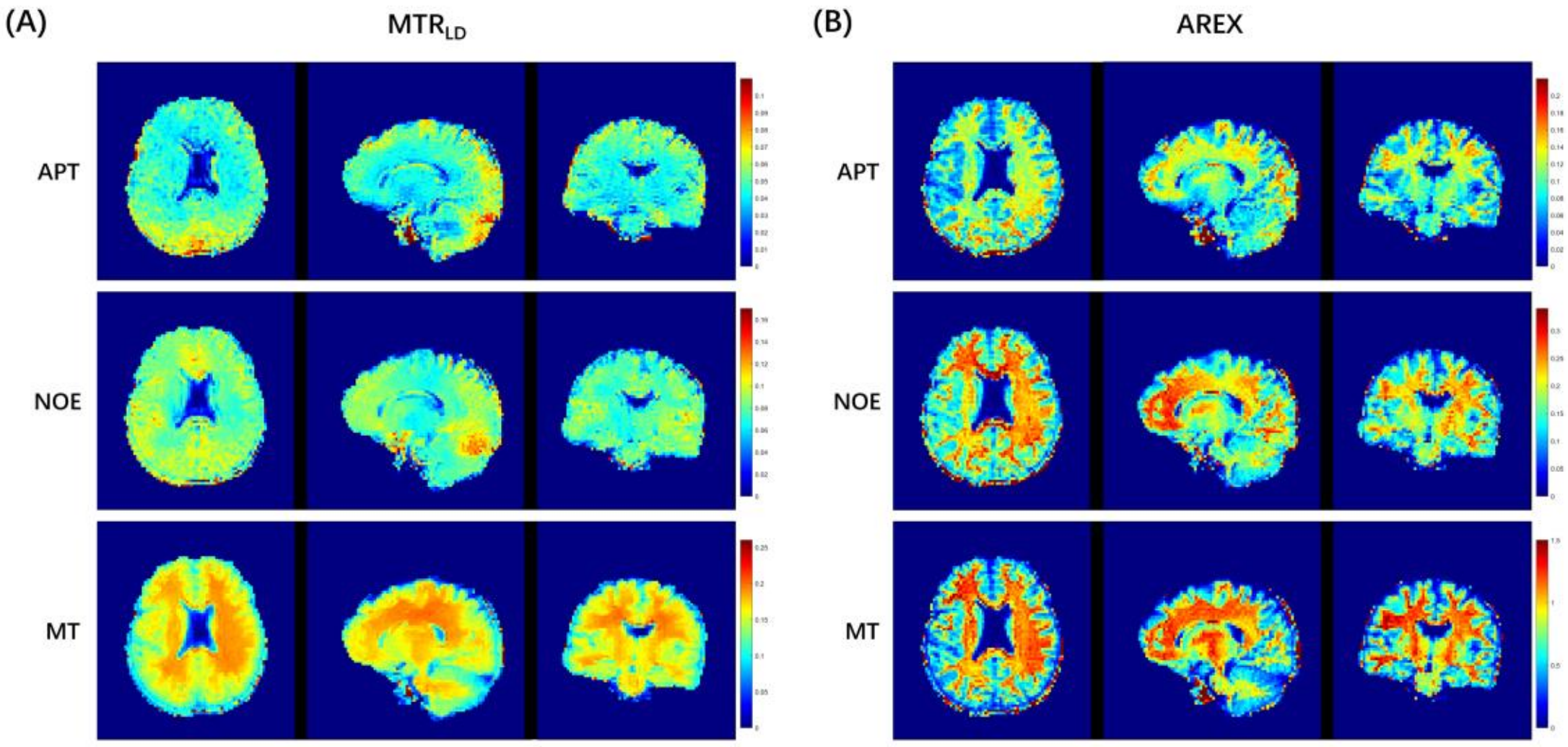

FIGURE 7 Comparison of final quantitative maps derived from $MTR_{LD}$ and AREX metrics of the second cohort. Representative whole-brain maps for amide proton transfer (APT), nuclear Overhauser enhancement (NOE), and magnetization transfer (MT) from a healthy subject are shown. (A) Quantitative maps calculated using the $MTR_{LD}$ metric. (B) Corresponding maps calculated using the AREX metric. All maps were generated using the complete processing pipeline, including neural network-based $B_1$ correction. $MTR_{LD}$, magnetization transfer ratio based on Lorentzian difference; AREX, apparent exchange-dependent relaxation; APT, amide proton transfer; NOE, nuclear Overhauser enhancement; MT, magnetization transfer.

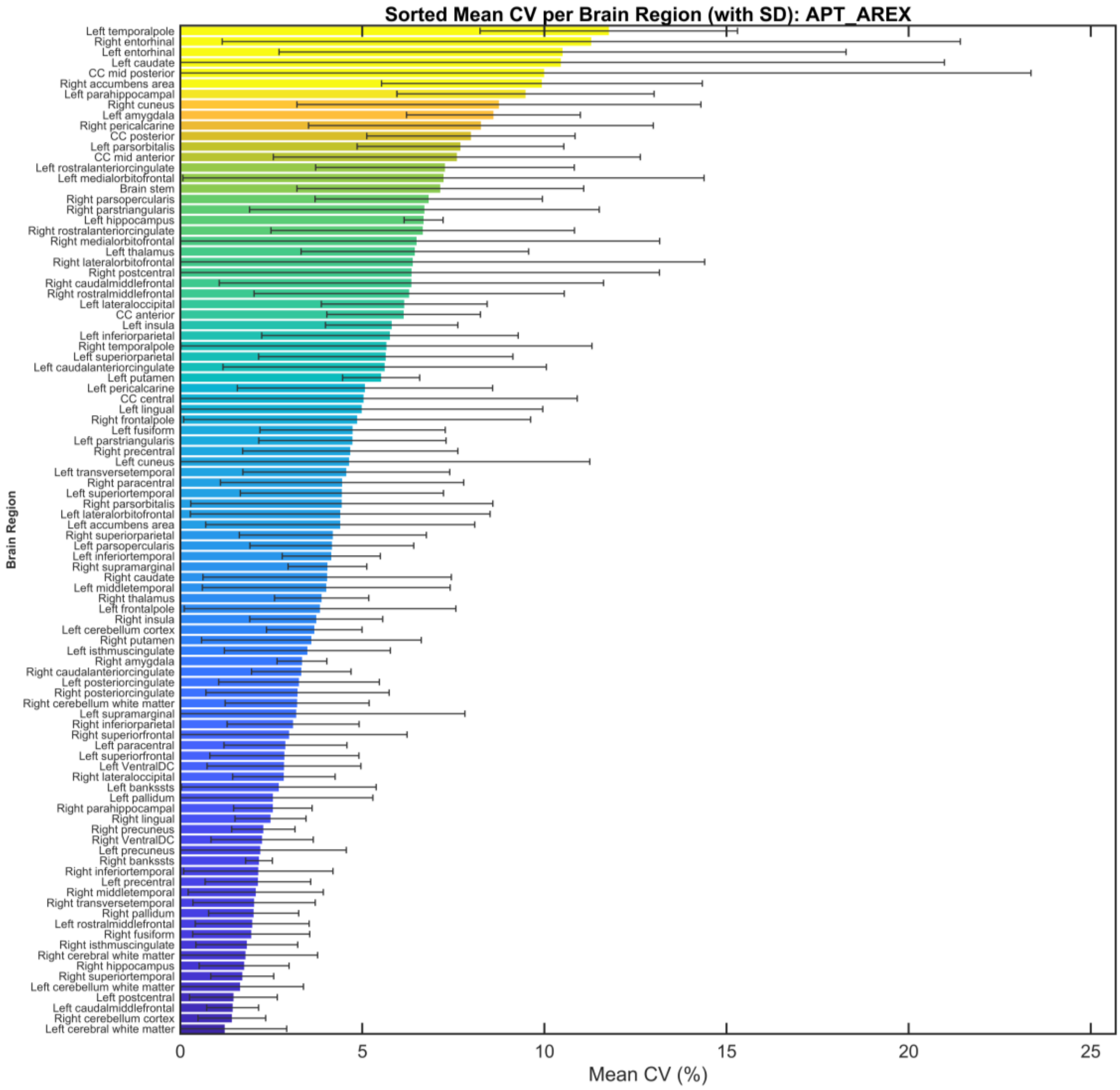

FIGURE 8 Test-retest reproducibility of the APT_AREX metric across human brain regions. The bar chart displays the mean coefficient of variation (CV) across 96 segmented brain regions, sorted in ascending order. Data were acquired from four healthy subjects, each scanned twice on alternate days. Each bar represents the mean CV across the four subjects, and the error bars indicate the standard deviation (SD). The analysis used the complete pipeline with neural network-based $B_1$ correction. Brain region abbreviations correspond to the FreeSurfer segmentation output (e.g., CC, Corpus Callosum). APT, amide proton transfer; AREX, apparent exchange-dependent relaxation; CV, coefficient of variation; SD, standard deviation.

## SUPPORTING INFORMATION

TABLE S1 MRI acquisition parameters for $B_0$, $B_1$, and $T_1$ mapping.

| parameter | Scan 1 | Scan 2 | Scan 3 | Scan 4 |
|---|---|---|---|---|
| **Sequence** | Spoiled GRE | Spoiled GRE | Spoiled GRE | Spoiled GRE |
| **Flip angle** | 135° | 260° | 4° | 16° |
| **TR (ms)** | 35 | 35 | 8.4 | 8.4 |
| **Number of echoes** | 1 | 1 | 2 | 2 |
| **TE (ms)** | 4.2 | 4.2 | 2.0/5.8 | 2.0/5.8 |
| **Bandwidth (Hz/pixel)** | 240 | 240 | 350 | 350 |
| **Field of view ($mm^3$)** | 220×220×180 | 220×220×180 | 220×220×180 | 220×220×180 |
| **Matrix size** | 96×96×72 | 96×96×72 | 128×128×106 | 128×128×106 |
| **Voxel size ($mm^3$)** | 2.3×2.3×2.5 | 2.3×2.3×2.5 | 1.7×1.7×1.7 | 1.7×1.7×1.7 |
| **Acceleration factor (GRAPPA)** | 4 | 4 | 2 | 2 |
| **Partial Fourier** | Phase 6/8 Slice 6/8 | Phase 6/8 Slice 6/8 | Slice 6/8 | Slice 6/8 |
| **Elliptical Sampling** | Yes | Yes | Yes | Yes |
| **Orientation** | sagittal | sagittal | sagittal | sagittal |
| **Scan time** | 45s | 45s | 43s | 43s |

TABLE S2 Initial values and boundaries for the center frequency (Δ), amplitude (A), and full width at half maximum (W) of the pools in the Lorentzian fitting model.

| **Z-spectrum contributor** | **Parameter** | **Starting value** | **Lower bound** | **Upper bound** |
|---|---|---|---|---|
| **Water** | Δ (ppm) | 0 | 0 | 0 |
| | A | 0.9 | 0.02 | 1 |
| | W (ppm) | 1.4 | 0.3 | 10 |
| **Amide** | Δ (ppm) | 3.5 | 3 | 4 |
| | A | 0.025 | 0 | 0.2 |
| | W (ppm) | 0.5 | 0.4 | 4 |
| **NOE** | Δ (ppm) | -3.5 | -4 | -3 |
| | A | 0.02 | 0 | 0.4 |
| | W (ppm) | 3 | 1 | 5 |
| **MT** | Δ (ppm) | -1 | -4 | 0 |
| | A | 0 | 0 | 1 |
| | W (ppm) | 25 | 15 | 100 |

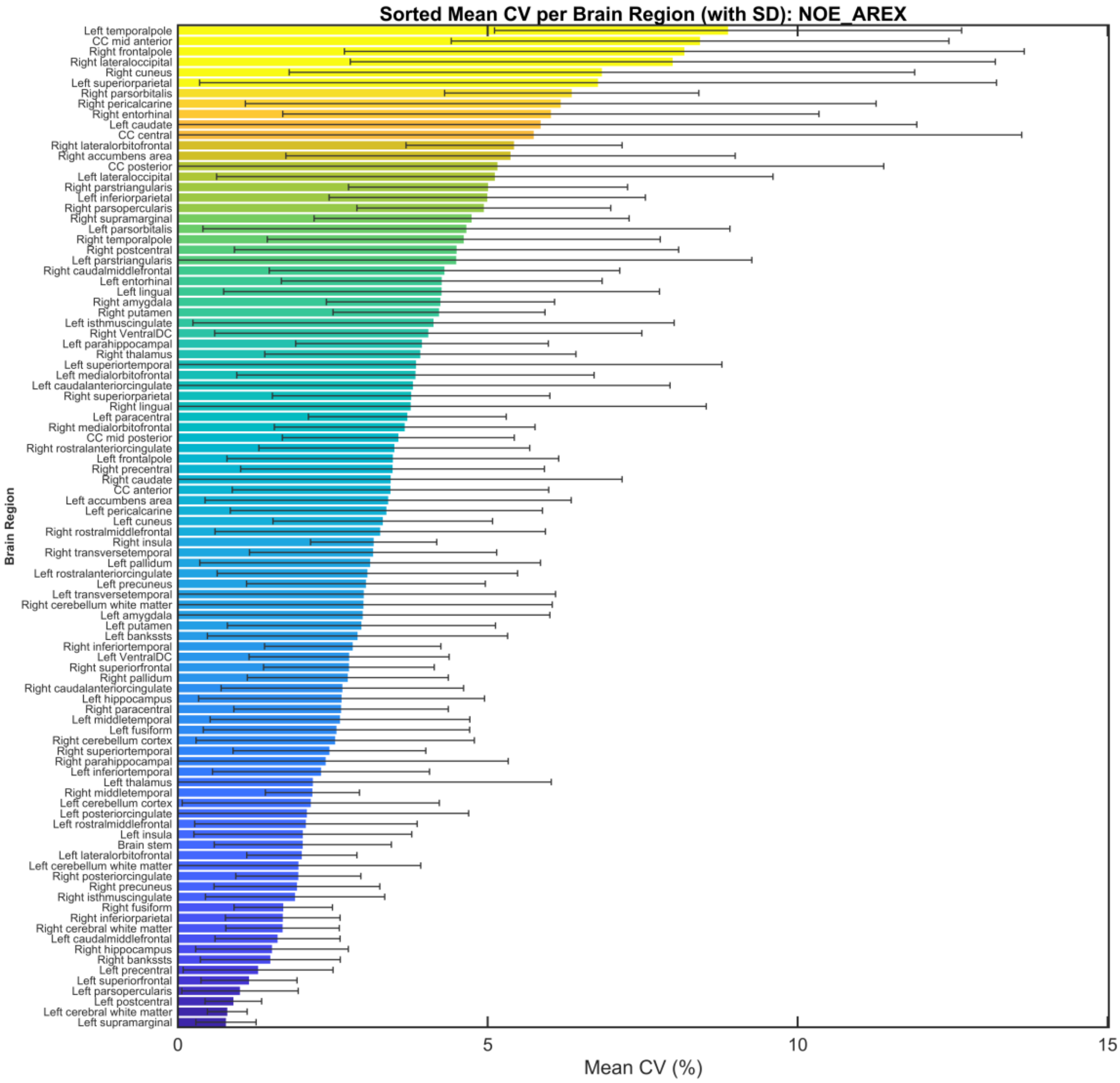

FIGURE S1 Test-retest reproducibility of the NOE_AREX metric across human brain regions. The bar chart displays the mean coefficient of variation (CV) across 96 segmented brain regions, sorted in ascending order. Data were acquired from four healthy subjects, each scanned twice on alternate days. Each bar represents the mean CV across the four subjects, and the error bars indicate the standard deviation (SD). The analysis used the complete pipeline with neural network-based $B_1$ correction. Brain region abbreviations correspond to the FreeSurfer segmentation output (e.g., CC, Corpus Callosum). NOE, nuclear Overhauser enhancement; AREX, apparent exchange-dependent relaxation; CV, coefficient of variation; SD, standard deviation.

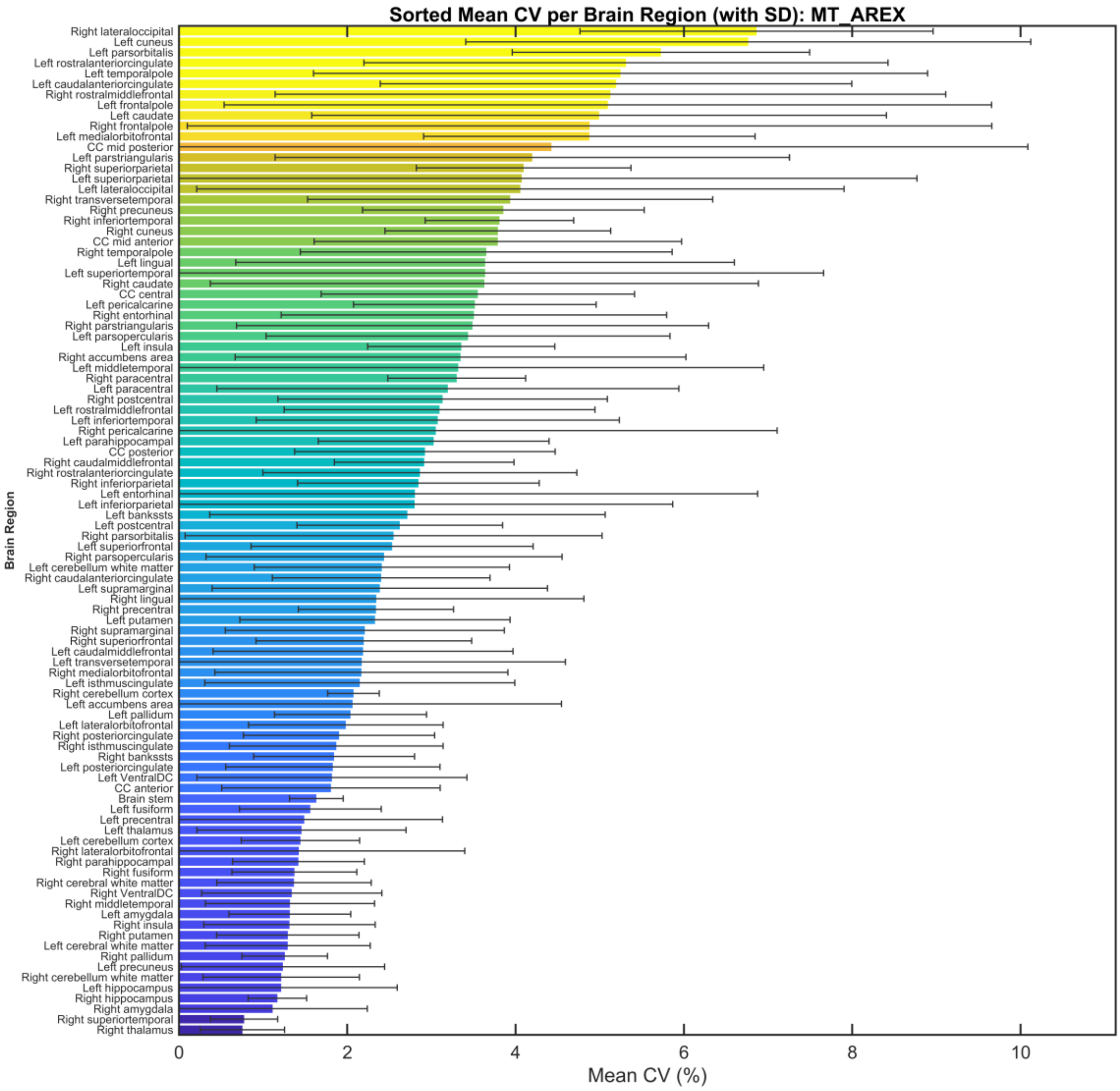

FIGURE S2 Test-retest reproducibility of the MT_AREX metric across human brain regions. The bar chart displays the mean coefficient of variation (CV) across 96 segmented brain regions, sorted in ascending order. Data were acquired from four healthy subjects, each scanned twice on alternate days. Each bar represents the mean CV across the four subjects, and the error bars indicate the standard deviation (SD). The analysis used the complete pipeline with neural network-based $B_1$ correction. Brain region abbreviations correspond to the FreeSurfer segmentation output (e.g., CC, Corpus Callosum). MT, magnetization transfer; AREX, apparent exchange-dependent relaxation; CV, coefficient of variation; SD, standard deviation.

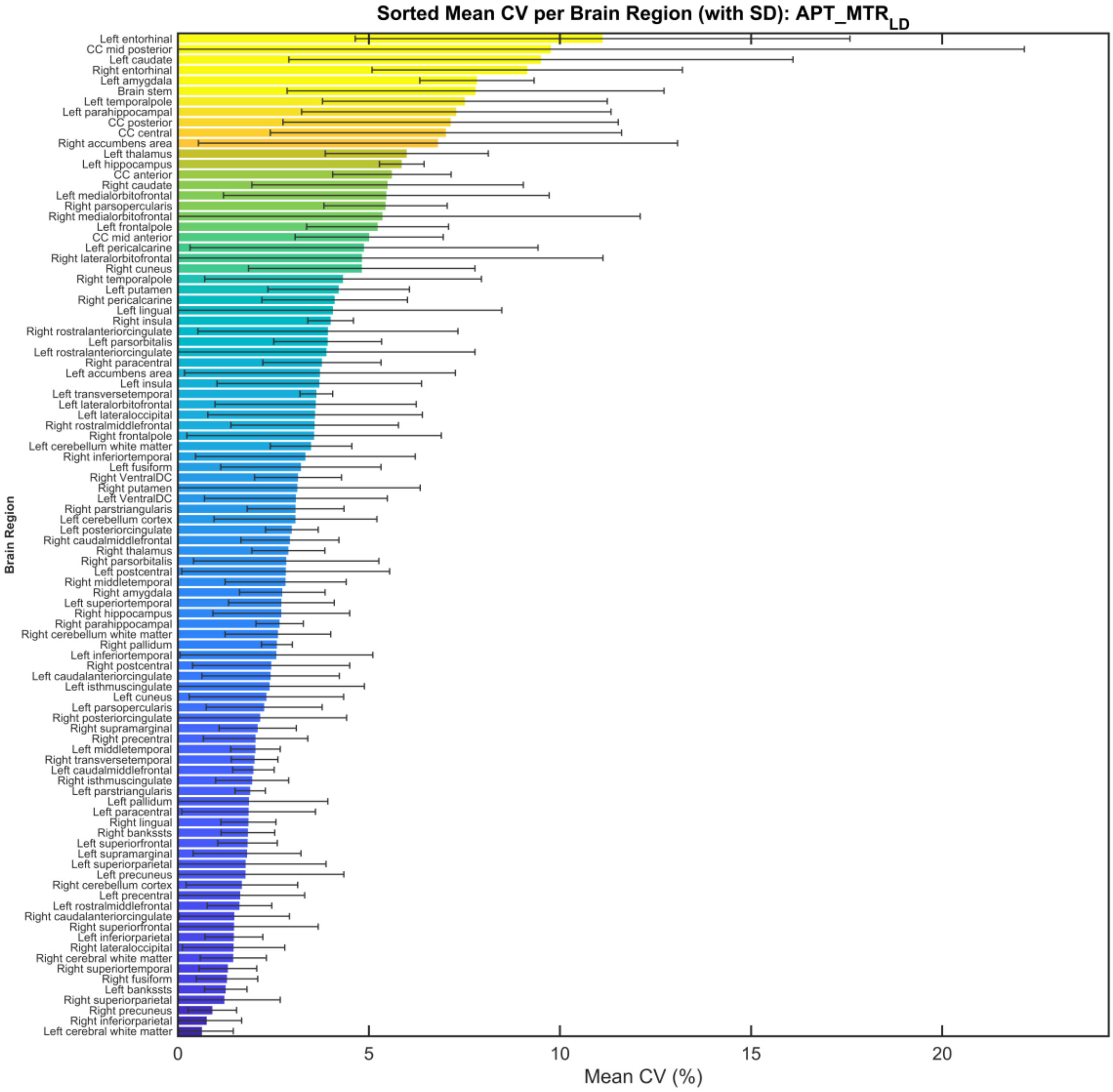

FIGURE S3 Test-retest reproducibility of the APT_MTR$_{LD}$ metric across human brain regions. The bar chart displays the mean coefficient of variation (CV) across 96 segmented brain regions, sorted in ascending order. Data were acquired from four healthy subjects, each scanned twice on alternate days. Each bar represents the mean CV across the four subjects, and the error bars indicate the standard deviation (SD). The analysis used the complete pipeline with neural network-based $B_1$ correction. Brain region abbreviations correspond to the FreeSurfer segmentation output (e.g., CC, Corpus Callosum). APT, amide proton transfer; MTR$_{LD}$, magnetization transfer ratio based on Lorentzian difference; CV, coefficient of variation; SD, standard deviation.

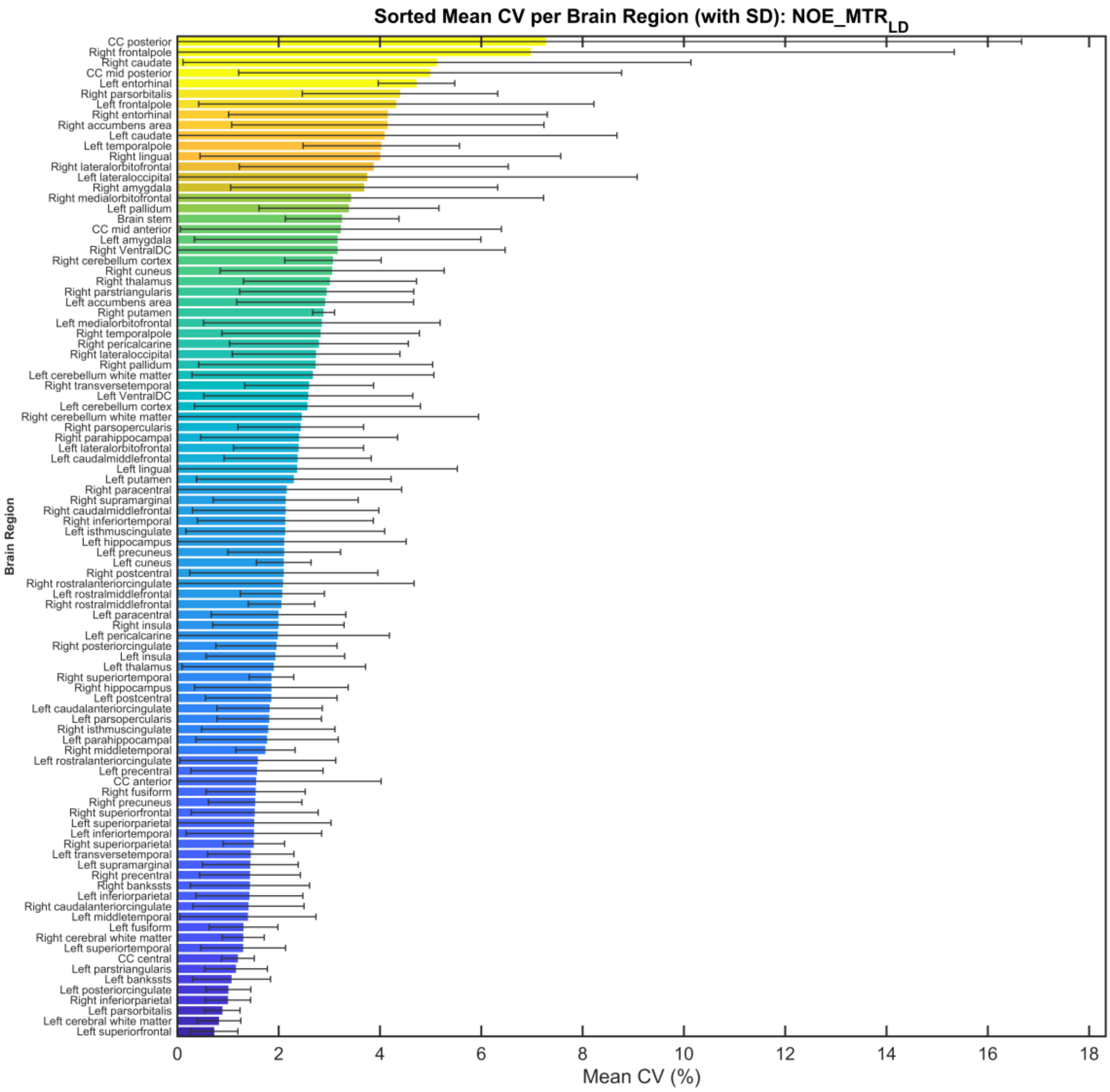

FIGURE S4 Test-retest reproducibility of the NOE_MTR$_{LD}$ metric across human brain regions. The bar chart displays the mean coefficient of variation (CV) across 96 segmented brain regions, sorted in ascending order. Data were acquired from four healthy subjects, each scanned twice on alternate days. Each bar represents the mean CV across the four subjects, and the error bars indicate the standard deviation (SD). The analysis used the complete pipeline with neural network-based $B_1$ correction. Brain region abbreviations correspond to the FreeSurfer segmentation output (e.g., CC, Corpus Callosum). NOE, nuclear Overhauser enhancement; MTR$_{LD}$, magnetization transfer ratio based on Lorentzian difference; CV, coefficient of variation; SD, standard deviation.

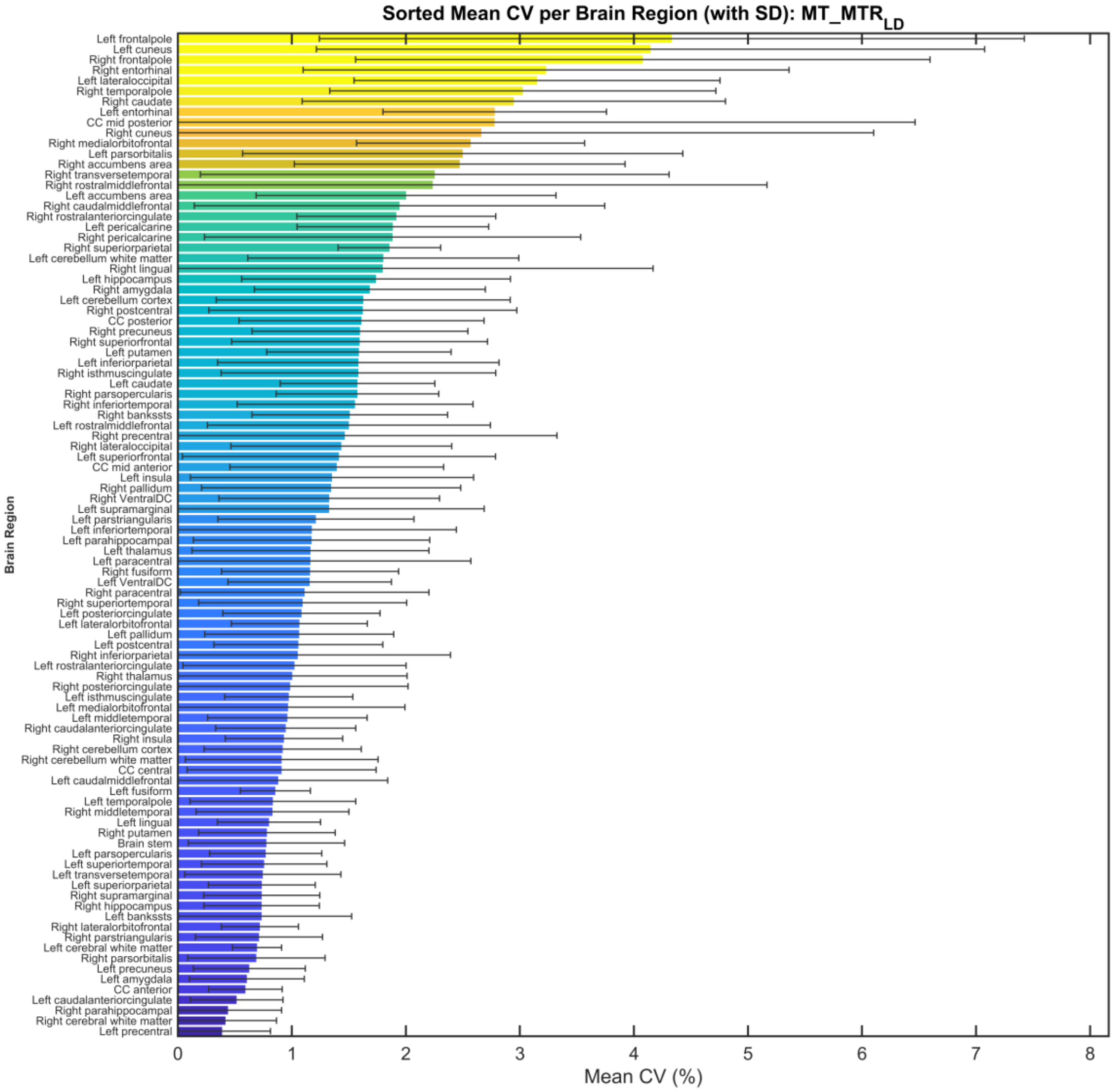

FIGURE S5 Test-retest reproducibility of the MT_MTR$_{LD}$ metric across human brain regions. The bar chart displays the mean coefficient of variation (CV) across 96 segmented brain regions, sorted in ascending order. Data were acquired from four healthy subjects, each scanned twice on alternate days. Each bar represents the mean CV across the four subjects, and the error bars indicate the standard deviation (SD). The analysis used the complete pipeline with neural network-based $B_1$ correction. Brain region abbreviations correspond to the FreeSurfer segmentation output (e.g., CC, Corpus Callosum). MT, magnetization transfer; MTR$_{LD}$, magnetization transfer ratio based on Lorentzian difference; CV, coefficient of variation; SD, standard deviation.